\documentclass[12pt]{article} 

\usepackage{amsmath}
\usepackage{amssymb, bbm, graphicx}

\title{ Digital Cavities and Their Potential Applications}

\author{Khadga Karki$^a$, Magne Torbj\"orsson$^a$, Julia R. Widom$^b$, Andrew H. Marcus$^b$,\\ and T\"onu Pullerits$^a$\\ 
\llap{$^a$} Chemical Physics, Lund University,\\
Getingev\"agen 60, 22241, Lund, Sweden\\
\llap{$^b$} University of Oregon,\\
Eugene, OR 97403-1253, USA\\
E-mail: Khadga.Karki@chemphys.lu.se}

\begin{document}

\maketitle

\begin{abstract}
The concept of a digital cavity is presented. The functionality of a tunable radio-frequency/microwave cavity with unrestricted Q-factor is implemented. The theoretical aspects of the cavity and its potential applications in high resolution spectroscopy and synchronization of clocks together with examples in signal processing and data acquisition are discussed.

\end{abstract}

\section{Introduction}
The development of analog to digital converters (ADC) is advancing rapidly. The cutting-edge ADCs available commercially have a bandwidth of 20 GHz and can digitize signals at astonishing speed of about 65 GS/s (giga samples per second). This rapid development in ADCs have enabled analysis of the signals from DC to microwave frequencies digitally. 

Traditionally resonator circuits (LC circuits) and cavity resonators have been used to analyze high frequency signals. The resonance frequency of an LC circuit can be tuned, which makes them useful in applications where the receiver needs the flexibility to analyze signals coming at different frequencies. However, due to the inherent power losses in the circuit the quality factor (Q-factor) that can be achieved in an LC circuit is only about $10^2$. Radio frequency cavities or microwave cavities, made up of a closed metal structure that confine electromagnetic field can have extremely high Q-factor (upto 10$^6$). Each cavity has a set of resonant modes (normal modes) that are determined by the physical parameters of the cavity. These are not easy to tune without compromising the performance of the cavity. Digital cavities overcome the limitation of these resonators, i.e. they are tunable, and at the same time, can have extremely high Q-factor. 

In the simple case of a linear cavity formed by two reflecting surfaces, the mode spacing or the free spectral range $fsr$ is given by
\begin{equation}\label{EQ1}
fsr = \frac{ c}{2 L},
\end{equation} 
where $c$ is the speed of light and $L$ is the distance between the surfaces. The frequencies of the normal modes are given by $N\cdot fsr$ with $N\in \{1,2,3,..\}$. A sinusoidal signal at the resonance frequency is amplified by the cavity due to constructive interference in the successive round trips in the cavity while the other signals are suppressed due to destructive interference. A time-periodic boundary condition applied to a sinusoidal signal leads to similar interference effects mimicking the functionality of a cavity. Here, we use the term time-periodic boundary condition to emphasize that the condition applied is similar to the periodic boundary condition (in spatial dimensions) used for example in molecular dynamics simulations ~\cite{LEACH, ALLEN, KARKI_2011B} to simulate properties of a large system by tessellation of a smaller simulation box. The periodic-boundary condition in time is implemented by translating the signal $x(t)$ at time $t$ to $x(t+k\cdot T)$, where $T$ is the period and $k\in \{0,1,2,...\}$. Implementation of such a boundary condition during the digitization of a signal by an ADC leads to the formation of a digital cavity.

\begin{figure}
\includegraphics[width=8.0cm]{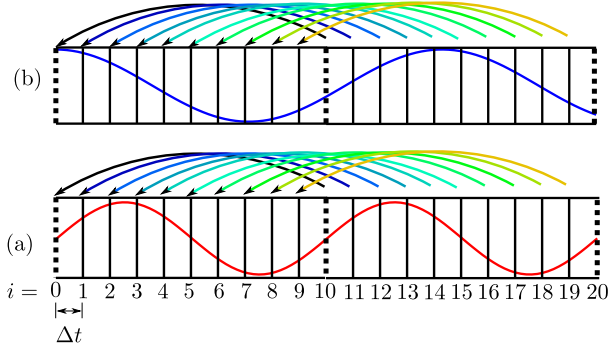}
 \caption{\label{FIG1} Digital cavity with  $n=10$ and cavity length $T=10 \Delta t$. In (a) the signal is resonant with the cavity while it is non-resonant in (b). }%
 \end{figure}

Fig.\ref{FIG1} shows a pictorial representation of a digital cavity implemented using discrete values of digitized signals. In the figure the signal is discretized at a fixed interval of $\Delta t$. A period of the digital cavity contains $n=10$ data points with period $T= n\cdot \Delta t$. Sinusoidal signals that are in resonance and off-resonance with the cavity are shown in panel (a) and (b), respectively.

\section{Theory}

A digital cavity of length $n$ is defined as 

\begin{equation}\label{EQ2}
y(j\cdot \Delta t;n):=\sum_{k=0}^{N_c} x(j\cdot \Delta t+k\cdot T); j=1,2,...n,
\end{equation}
where the waveform $y$ is the response of the cavity to the signal $x$ and $N_c+1$ is the number of times the cavity function is applied to the signal. The cavity defined by Equ.\eqref{EQ2} amplifies the resonant frequencies while suppressing (or filtering out) the non-resonant frequencies. 

\begin{figure}
\includegraphics[width=8.0cm]{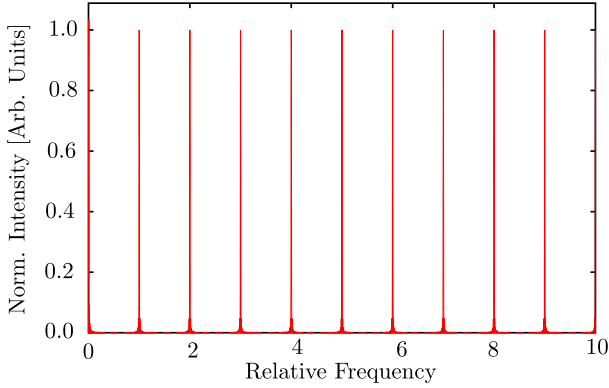}
 \caption{\label{FIG2} A frequency comb that is amplified by a digital cavity. The $x$-axis is presented as frequency coordinate relative to the fundamental mode. The amplification as well as the line width of the resonant modes depend on $N_c$. }%
 \end{figure}

 The fundamental mode of the cavity is given by $f_0=1/(n\cdot \Delta t)$, which can be tuned either by changing $n$ or $\Delta t$. In an analogous way a physical cavity can be tuned either by changing the cavity length or the dielectric constant of the medium in the cavity. The free spectral range of a digital cavity is the same as the fundamental mode ($fsr = f_0$), hence the cavity amplifies a frequency comb~\cite{CUNDIFF2005} with frequencies that are integer multiple of the fundamental mode, i.e. $f_0, 2f_0, 3f_0,...$ including the DC signal as shown in Fig.\ref{FIG2}. We denote the resonant frequencies by $f_{res}$. Note that a DC signal is also amplified by an analog cavity. The DC offset of a signal in a digital cavity can be filtered out using a high band pass filter or an AC coupling prior to digitization. Another simple approach to filter out the DC offset is to use a generalized digital cavity as defined later. 

\begin{figure}
\includegraphics[width=7.5cm]{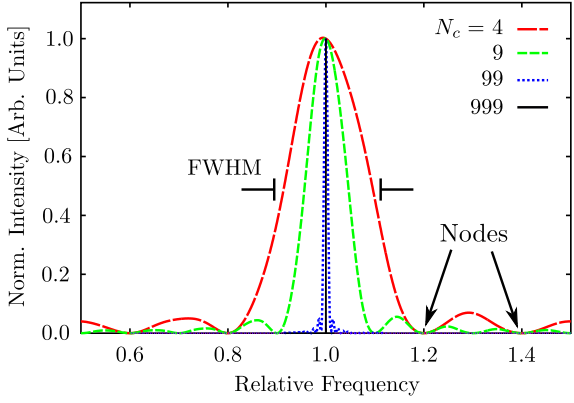}
 \caption{\label{FIG3} Line-width of a digital cavity using $r=2$ and different values of $N_c$. The line-width narrows progressively with increasing $N_c$. The figure also shows that at certain non-resonant frequencies, indicated as nodes, the cavity completely suppresses the signal. }%
 \end{figure}

The upper limit of the response of the cavity to a non-resonant frequency $f$ is given by
\begin{equation}\label{EQ3}
||y(j\cdot \Delta t)||\leq \frac{\sqrt{5}}{2\pi}\frac{f_{res}}{f-f_{res}},
\end{equation}
where $f_{res}$ is the resonant frequency closest to $f$ (see Appendix for the derivation), and the norm in the left side of the equation is taken to be the absolute value of the waveform. As shown by Equ.\eqref{EQ3} there is no upper bound for resonant frequencies. In this case the waveform gets amplified linearly with $N_c$, i.e. $||y(j\cdot \Delta t)|| = (N_c+1) ||x(j\cdot \Delta t)||$, and the cavity response can be made arbitrarily large by choosing $N_c$ appropriately. If we define the amplitude of the response of the cavity to a signal as $A(f)=max(||y(j\cdot \Delta t)||)$ then the ratio of the amplitude for a resonant frequency to a non-resonant frequency, $A(f_{res})/A(f)$, can be arbitrarily large. This, in other words, means that the line-width of the cavity can be arbitrarily small. We take full-width at half the maximum (FWHM) as the measure of the line-width. Let $\xi_{f_0}$ denote the half-width at half the maximum. Then for $f = f_0+ \xi_{f_0}$ we have $A(f) = (N_c+1)/2$, and using Equ.\eqref{EQ3} we get
\begin{equation}\label{EQ4}
2\xi_{f_0} = \frac{2 \sqrt{5} }{\pi(N_c+1)}.
\end{equation} 
As the waveform generated after the cavity, $y(j\cdot \Delta t)$, is a set of numbers, the definition of the amplitude can be generalized. If the amplitude is defined as $A(f) = max(||(y(j\cdot \Delta))^r||)$, where $r \geq 1$, then the generalized FWHM is given by  
\begin{equation}\label{EQ5}
2\xi_{f_0} = \frac{2^{1/r}\sqrt{5} }{\pi (N_c+1)}.
\end{equation}
As shown by Equ.\eqref{EQ5} the line-width of a digital cavity can be made arbitrarily narrow by choosing large $N_c$, and in addition it can also be digitally narrowed by simply choosing larger $r$. Fig.\ref{FIG3} shows line-widths for a digital cavity with different values of $N_c$ and $r=2$.

Fig.\ref{FIG3} also shows nodes at certain frequencies at which a digital cavity completely suppresses the signal even for low values of $N_c$. The frequencies at which these nodes occur depend on $N_c$, and they occur at $f=f_{res}+ p f_0/(N_c+1)$ for $p\in \{1,2,...,N_c\}$. As pointed out later these nodes can be useful in demultiplexing a signal that is multiplexed over different frequencies.

\section{Discussion on the Applications and Scope}
\begin{figure}
\includegraphics[width=7.5cm]{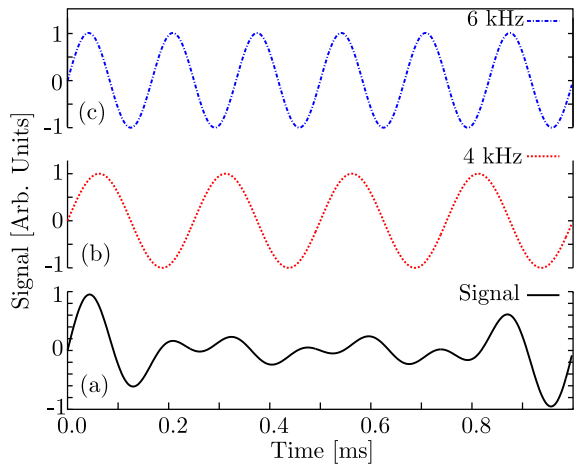}
 \caption{\label{FIG4} Signal de-multiplexing using a digital cavity. Panel (a) shows the signal that is generated using direct digital synthesis by summing signals at frequencies 4, 5, 6 and 7 kHz. All the frequencies have same amplitude and zero phase offset. The signal in (a) is digitized using an ADC at a rate of 96 kS/s. A digital cavity with $n=24$ selects the signal at 4 kHz (panel (b)) and another cavity with $n=16$ selects the signal at 6 kHz (panel(c)). $N_c$ is set to 3 and 5 in (b) and (c), respectively. All the waveforms have been normalized by the maximum value. }%
 \end{figure}

One of the scientific applications of a digital cavity is in phase-sensitive signal detection. We developed the concept for real time analysis of phase and amplitude of different frequencies present in the fluorescence signal from a sample that is excited with phase modulated laser pulses as in the fluorescence detected wave-packet interferometry~\cite{RICE1991,MARCUS2006} and the two-dimensional electronic coherence spectroscopy.~\cite{MARCUS2007} In these experiments, the fluorescence from the sample is multiplexed over many frequencies due to the contribution of the linear and the non-linear response of the sample. The signals at different frequencies provide a plethora of information about the system being investigated. A data-acquisition system that is capable of de-multiplexing the different contributions simultaneously would make this technique promising for chemical imaging. This in conjunction with near-field scanning techniques, can provide detailed information of chemical composition of a system with sub-wavelength resolution.~\cite{KARKI_2012B} A digital cavity can be used to de-multiplex the signal and analyze the different contributions simultaneously. Thus, it could enable multiplexed chemical imaging using the fluorescence detected interferometric techniques. Fig.\ref{FIG4} shows two signals at 4 and 6 kHz (panels (b) and (c), respectively) recovered from a signal that is synthesized by digitally mixing four frequencies at 4, 5, 6 and 7 kHz. The signal is digitized at the rate of 96 kS/s (kilo samples per second). Two digital cavities with $n=24$ and 16 are implemented in real time to recover the two different contributions to the signal. Other contributions can also be recovered at the same time simply by implementing other cavities with different cavity lengths. 

De-multiplexing technique as shown in Fig.\ref{FIG4} can also be used in data transmission employing multiple frequencies. When the frequencies are well defined, even small values of $N_c$ can select the signal contribution from the desired frequency.  The 4 kHz signal in Fig.\ref{FIG4} is selected by using $N_c=3$ and the 6 kHz signal is selected by using $N_c=5$. Note that for the 4 kHz signal all the other frequencies, 5,6 7 kHz, lie on the nodes of the cavity response when $N_c=3$, i.e. they are completely suppressed, and the same is true for the 6 kHz contribution when $N_c=5$. Multiplexed data transmission can dramatically improve the data transfer rate compared to the transmission using single frequency as is commonly done in wireless communication. Clever implementation of digital cavities can have significant impact in this field.

A digital cavity described by Equ.\eqref{EQ2} is generalization of coherent sampling technique~\cite{NEUVO1984,SVANBERG2007} in which a common reference clock is used for signal generation as well as for digitization. Coherent sampling has been used for phase-sensitive detection, similar to that shown in Fig.\ref{FIG4}, of signals mainly at $f_0$ and $2f_0$ to investigate rare gases using high-resolution tunable diode laser spectroscopy.~\cite{SVANBERG2007} A digital cavity is generalization of the previously implemented techniques in that it does not need the same reference clock for signal generation and digitization. Furthermore the definition of the digital cavity can be generalized as 
\begin{align}\label{EQ6}
y(j\cdot \Delta t):=&\sum_{k=0}^{N_c} x(j\cdot\Delta t+n\cdot \Delta t\cdot k) \cos(k\cdot\theta)+\nonumber\\
&\sum_{k=0}^{N_c} x((j+l)\cdot\Delta t+n\cdot \Delta t\cdot k) \sin(k\cdot\theta),
\end{align} 
where $l\in \{n/4,n/2,1,2,3,..\}$ is used to offset the fundamental frequency $f_0$ to $f_g = nf_0/(4l)$, and $\theta = 2 \pi k(4l-n)/(4l)$ (see Appendix for the derivation). Equ.\eqref{EQ2} is a special case of the generalized cavity in which $4l=n$ . For $4l=2n$ the generalized cavity functions as a comb-pass filter amplifying signals at $f_0/2, 3f_0/2, 5f_0/2,...$, and for other $l$ it acts as a comb-pass filter amplifying signals at $f_g, f_g+f_0, f_g+2f_0,...$. Note that a DC offset of a signal in the generalized cavity, except for the case when $4l=n$, becomes non-resonant with the cavity. The generalized digital cavity can be tuned by changing either of $\Delta t$, $n$ or $l$. Such tunability significantly increases the scope of a digital cavity compared to previously known techniques like the coherent sampling.

\begin{figure}[htbp]
\includegraphics[width=8cm]{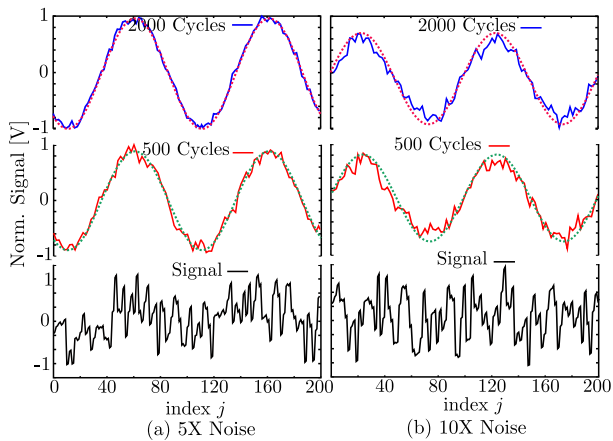}
 \caption{\label{FIG5} Reduction of random (white) noise in a digital cavity due to waveform averaging. The signals are generated by mixing random noise five and ten times the amplitude  of a sine function at 960 Hz in (a) and (b), respectively. Digitization is done at the rate of 96 kS/s. $x-$ axis shows the index $j$ used to denote the digitization position - one cycle is completed for $j=100$; two cycles are shown in the figures. The signal to noise ratio improves in both the cases ((a) and (b)) with larger $N_c$. The dotted lines are fits to the cavity response using sine functions. }%
 \end{figure}

As a digital cavity is freely tunable to signal source(s) at a particular frequency it can be used in applications like software defined radio in which the tuning of the radio to a certain frequency is done by the software.  When a signal needs to be transmitted over a noisy channel then the cavity can be used to suppress the random (white) noise. Note that the cavity averages a cycle of the resonant signal $N_c$ times, which improves the signal to noise ratio by a factor of $\sqrt{N_c}$ for the random noise.  Fig.\ref{FIG5} shows two sinusoidal signals retrieved from their corresponding noisy parent signals. Some examples where suppression of white noise becomes important is in signal transmission over long distance (satellite communication, interplanetary communication, etc.),  detection of faint radio/microwave frequencies from astronomical bodies, detection of cosmic microwave background, and perhaps even signals from extra-terrestrial intelligent life forms as sought after by the SETI project. In some cases deliberate adulteration of signal with white noise can be useful to avoid eavesdropping. Digital cavities can significantly improve the instrumentation and reduce the cost related to high-end data acquisition/analysis systems needed in the works mentioned above.  

Another potentially important application of the digital cavity can be in high resolution spectroscopy. As shown in Equ.\eqref{EQ5} the line-width of the cavity becomes narrower with increasing $N_c$, i.e. data acquisition time. This is the consequence of the fact that the precision of any interferometer depends on the number of the cycles of the sinusoidal signal that interfere.~\cite{VAUGHAN1977} In analog cavities the multi-pass within the cavity increases the number of cycles that interfere, however, the power losses in each pass sets an upper limit. Power loss in this case equates to information loss. A digital cavity is not prone to this limitation as information present in each cycle is stored digitally. Consequently, a digital cavity can have arbitrarily high precision, depending on the time over which the data is acquired. Data acquired for 1s gives a precision of about 1 Hz in the measurement of a frequency, and higher measurement time results in better precision regardless of the frequency being measured. Every digitizer, however, has the problem of electronic jitter, which introduces uncertainty in time at which digitizations happen. Jitter in digital cavity is analogous to various noise sources in an analog cavity like the thermal noise and zero point fluctuation (vacuum fluctuations). In the high end digitizers, jitter can be less than 100 fs. The jitter does not set an upper limit but increases the data acquisition time to reach the desired precision. With the currently available digitizers it would be possible to analyze the hyperfine transitions in atoms and molecules in general, and atoms used for atomic clocks in particular, with extremely high precision. As the digitization is controlled by a clock, precision measurements with a digital cavity can also be used to synchronize clocks. When a microwave of 20 GHz generated by a source controlled by a master clock is used for synchronization then the frequency measurement done by a digital cavity can be used to synchronize the clock of the digitizer to $\approx 10^{-14}$s, and a day of measurement enables synchronization better than $10^{-15}$s , which rivals the synchronization that can be achieved using analog techniques based on radio/microwave freqeuncies.~\cite{CUNDIFF2005} However,implementing a digital cavity with stable operation for a day or longer could be challenging.

\section{Conclusion}
We have presented the theoretical description of the digital cavity and derived some of its properties. Digital cavities, in general, implement the functionality of the analog cavities in time domain and eliminate problems related to power losses in the analog cavities. We have also shown examples of how digital cavities can be used in signal analysis and discussed other possible uses in signal transmission, high precision spectroscopy and precise synchronization of clocks.

\textbf{Acknowledgments}

KK thanks Werner-Gren Foundation for the generous post-doctoral fellowship awarded to him.
Financial support from the Knut and Alice Wallenberg Foundation is greatfully acknowledged.

\vspace{1cm}
\section{Appendix A:} \textbf{ Bounded and unboundedness of the cavity response}

To simplify the derivation of the properties of a digital cavity we represent the signal $x ( i \cdot \Delta t)$, $i \in \mathbbm{Z}$ as a Fourier series. Without any loss of generality, we consider a component of the series, a sine function, $\sin ( \omega t + \phi)$, digitized at time intervals $\Delta t$. We use the following equation to to define a digital cavity of length $n$, $n \in \mathbbm{N}^+$.
\begin{equation}
  y ( j \cdot \Delta t ; n) : =  \sum^{N_c}_{k = 0} \sin ( \omega \cdot j
  \cdot \Delta t + \omega \cdot k \cdot n \cdot \Delta t + \phi) . 
  \label{AEQ3}
\end{equation}
In (\ref{AEQ3}) $n \cdot \Delta t = T$ gives the period of the cavity, thus for each $j \in ( 0, 1, \ldots, n - 1)$ the equation sums up the values at equivalent points in the successive periods for $N_c$ times. Instead of using frequency $f$ as done in the main text, we use angular frequency $\omega$ in the equation to reduce notations in the equations. We define $\omega_c = 2 \pi / T$ as the fundamental frequency of the cavity. Using this definition, Equ.(\ref{AEQ3}) can be written as:
\begin{equation}
  y ( j \cdot \Delta t)  =  \sum_{k = 0}^{N_c} \sin \left( \omega \cdot j
  \cdot \Delta t + \frac{2 \pi \omega}{\omega_c} k + \phi \right) . 
  \label{AEQ4}
\end{equation}

We further analyse Equ.(\ref{AEQ4}) in the following two cases:

First, when $\omega = \omega_c$, i.e. when the cavity is resonant with the
function, the cavity simply amplifies the sine function $N_c + 1$ times: \ \
\begin{equation}
  y ( j \cdot \Delta t)  =  ( N_c + 1) \sin ( \omega_c \cdot j \cdot \Delta
  t + \phi) .  \label{AEQ5}
\end{equation}

Equ.(\ref{AEQ5}) is true for any $\omega = m \omega_c$ , where $m \in
\mathbbm{Z}$. Thus for all the frequencies that are integer multiple of the cavity frequency, application of the digital cavity amplifies the signal and at the same time increases the signal to noise ratio. As there are no any physical constrains that limit the amplification, it does not saturate. In other words, the amplification factor is not bounded.

Now we analyze the case when $\omega \neq m \omega_c$, i.e. when the cavity is not resonant with the function. For simplicity we analyse the case $\omega_c \leqslant \omega \leqslant 1.5 \omega_c$. By symmetry, the following argument also applies for the other cases. We can rewrite Equ.(\ref{AEQ4}) as
\begin{equation}
  y ( j \cdot \Delta t)  =  \sum_{k = 0}^{N_c} \sin \left( \frac{2 \pi
  \omega}{\omega_c} k + \phi_2 \right),  \label{AEQ6}
\end{equation}
where $\phi_2 = \omega \cdot j \cdot \Delta t + \phi$. Let $\omega / \omega_c
= 1 + x$ with $0 \leqslant x \leqslant 0.5$. Equ.(\ref{AEQ6}) can be written as
\begin{equation*}
  y ( j \cdot \Delta t)  =  \cos \phi_2 \sum_{k = 0}^{N_c} \sin ( \Omega k) + \sin \phi_2   \sum_{k = 0}^{N_c} \cos ( \Omega k), \;\; \textrm{with}\;\; \Omega = 2 \pi x.
\end{equation*}
The sums on the right hand side have an upper bound, which can be calculated as follows. Assume $N_c$ is very large. Take $r$ such that $\Omega r \leqslant
\pi \leqslant \Omega ( r + 1)$. Then
\begin{equation}
  \left\| \sum_{k = 0}^{N_c} \sin ( \Omega k) \right\|  \leqslant \sum_{k = 0}^r \sin ( \Omega k) \frac{\Delta k}{\Delta k} .
\end{equation}
Let $\Omega k = y$, then $\Delta k = \frac{\Delta y}{\Omega}$. Using $\Delta
k$ in the above
\begin{eqnarray*}
  \left\| \sum_{k = 0}^{N_c} \sin ( \Omega k) \right\| & \leqslant &
  \frac{\sum_{y = 0}^{\pi} \sin y \Delta y}{\Omega \Delta k}\\
  & \leqslant & \frac{\int_{y = 0}^{\pi} \sin y d y}{\Omega} = \frac{\omega}{\pi ( \omega - \omega_c)}
  .
\end{eqnarray*}

Similarly
\begin{equation*}
  \left\| \sum_{k = 0}^{N_c} \cos ( \Omega k) \right\|  \leqslant 
  \frac{\omega_c}{2 \pi ( \omega - \omega_c)} .
\end{equation*}
Thus for $\omega \neq \omega_c$, $\mathcal{F} ( j \cdot \Delta t)$ is bounded by
\begin{equation}
  \| \mathcal{F} ( j \cdot \Delta t) \|  \leqslant  \frac{\omega_c}{2 \pi (
  \omega - \omega_c)}  ( 2 \cos \phi_2 +  \sin \phi_2) . 
  \label{AEQ7}
\end{equation}
Note that the singularity at $\omega_c$ indicates the divergence of digital cavity map at $\omega_c$, which is also clear in Eq.(\ref{AEQ5}).

The maximum value \ $( 2 \cos \phi_2 + \sin \phi_2)$ can have is $\sqrt{5}$, which gives
\begin{equation}
  \| \mathcal{F} ( j \cdot \Delta t) \|  \leqslant  \frac{\sqrt{5}}{2 \pi} 
  \frac{\omega_c}{\omega - \omega_c} .  \label{AEQ8}
\end{equation}
Equ.(\ref{AEQ8}) shows that for the case when $\omega \neq \omega_c$, or in general when $\omega \neq m \omega_c$, the cavity response has an upper bound.

{\textbf{Determination of frequencies completely suppressed by a digital
cavity.}}

If we take $\omega = \omega_c / N_c$ in Equ.(\ref{AEQ6}) then
\begin{eqnarray}
  y ( j \cdot \Delta t) & = & \sum_{k = 0}^{N_c} \sin \left( \frac{2 \pi
  k}{N_c} + \phi_2 \right) = 0  \label{ANSUP}
\end{eqnarray}
for all $j$ as the summation is over a period of the sine function. Thus for a
digital cavity in which the successive periods are summed $N_c + 1$ times
there are $N_c$ frequencies between 0 and $\omega_c$ given by $\omega = p
\omega_c / ( N_c + 1)$ where $p \in \mathbbm{Z} \textrm{and} p \leqslant N_c$
that are completely filtered out. We define $\omega_c / N_c$ as the node
distance. \ The nodes are present at $\omega = i \cdot \omega_c \pm p \omega_c
/ ( N_c + 1)$ for all $i \in \mathbbm{N} \wedge p \in \{ 1, 2, \ldots, N_c
\}$.

\section{Appendix B:}

{\textbf{Frequencies selected by digital cavity and tuning the cavity.}}

According to Equ.(\ref{AEQ4}) for all $\omega = i \omega_c$ where $i$ is an
integer we have
\begin{equation}
  y ( j \cdot \Delta t)  =  ( N_c + 1) \sin ( \omega \cdot j \cdot \Delta t + \phi) . 
  \label{AEQ12}
\end{equation}
Thus the cavity selects all the frequencies that are integer multiple of the
fundamental frequency, in other words it selects a frequency comb with carrier
offset frequency and comb tooth spacing both given by $\omega_c$. The
fundamental frequency, and thereby the comb, selected by the cavity can be
changed either by changing $n$ (number of samples digitized per unit time) or
$\Delta t$ (digitization interval); remember $\omega_c = 2 \pi / ( n \cdot
\Delta t)$. However, the definition of the digital cavity can be generalized
such that carrier offset frequency and comb tooth spacing are independent. For
e.g. if we define the cavity as
\begin{eqnarray}
  y ( j \cdot \Delta t) & := & \sum_{k = 0}^{N_c} ( - 1)^k \sin \left(
  \omega \cdot j \cdot \Delta t + \frac{2 \pi \omega}{\omega_c} k + \phi
  \right)  \label{AEQ13}
\end{eqnarray}
then the carrier offset frequency is given by $\omega_c / 2$ while the tooth
spacing is given by $\omega_c$. Equ (\ref{AEQ13}) is a special case of the
general digital cavity defined as
\begin{equation}
  y ( j \cdot \Delta t ; \theta) := \sum_{k = 0}^{N_c} \sin \left( \omega \cdot j \cdot \Delta t + 2 \pi
  \left( \frac{\omega + \frac{\theta \omega_c}{2 \pi}}{\omega_c} \right) k +
  \phi \right) . \label{AEQ14}
\end{equation}
For $\omega$ to be resonant with the cavity we need
\begin{eqnarray}
  \frac{\omega + \frac{\theta \omega_c}{2 \pi}}{\omega_c} & = & 1 \nonumber\\
  \Rightarrow   \theta & = & 2 \pi \left( 1 - \frac{\omega}{\omega_c} \right)  \label{AEQ16}
\end{eqnarray}
The values $\theta$ can have is contrained due to the fact that data is
accumulated at discrete time points and $\phi$ is not known in general. The
general expression for the allowed values of $\theta$ in a digital cavity can
be found as follows:

Equ(\ref{AEQ14}) can be written as
\begin{equation*}
  y ( j \cdot \Delta t)  =  \sum_{k = 0}^{N_c} \sin \left( \omega \cdot j \cdot \Delta t +
  \frac{2 \pi \omega}{\omega_c} k + \phi \right) \cos ( \theta k) + \sum_{k =
  0}^{N_c} \sin \left( \omega \cdot j \cdot \Delta t + \frac{2 \pi
  \omega}{\omega_c} k + \phi + \frac{\pi}{2} \right) \sin ( \theta k).
\end{equation*}
Using the relations
\begin{equation*}
  \omega_c  =  \frac{2 \pi}{n \cdot \Delta t}, 
\end{equation*}
and
\begin{equation*}
  \frac{\pi}{2}  =  \frac{\omega_c \cdot n \cdot \Delta t}{4}
\end{equation*}
we analyze the second term
\begin{equation*}
  2 \textrm{nd}\; \textrm{term}  =  \sum_{k = 0}^{N_c} \sin \left( \omega \cdot j
  \cdot \Delta t + \omega \cdot n \cdot \Delta t \cdot k + \frac{n}{4} \cdot
  \Delta t \cdot \omega_c + \phi \right) .
\end{equation*}
Let $\omega_c = r \omega$, then we have
\begin{equation*}
  2 \textrm{nd}\; \textrm{term}  =  \sum_{k = 0}^{N_c} \sin \left( \omega \cdot j
  \cdot \Delta t + \omega \cdot n \cdot \Delta t \cdot k + \frac{n}{4} \cdot
  \Delta t \cdot r \cdot \omega + \phi \right)
\end{equation*}
According to Equ.(\ref{AEQ16}) we also have
\begin{equation}
  \theta  =  2 \pi \left( 1 - \frac{1}{r} \right)  \label{AEQ17}
\end{equation}
Let
\begin{eqnarray*}
  \frac{n \cdot r}{4} & = & l, \textrm{with}\; l \in \mathbbm{Z} \;
  \textrm{then}\\
  r & = & \frac{4 l}{n},\; \textrm{and}\\
  \theta & = & 2 \pi \left( \frac{4 l - n}{4 l} \right) .
\end{eqnarray*}
The frequency that resonates with the cavity is then given by
\begin{equation}
  \omega  =  \frac{n \omega_c}{4 l} .  \label{AEQ18}
\end{equation}
Finally the general expression for the cavity can be written as
\begin{eqnarray}
  y ( j \cdot \Delta t) 
  & = & \sum_{k = 0}^{N_c} \sin ( \omega \cdot j \cdot \Delta t + \omega
  \cdot n \cdot \Delta t \cdot k + \phi) \cos \left( 2 \pi \left( \frac{4 l -
  n}{4 l} \right) k \right) +  \nonumber\\
  &  & \sum_{k = 0}^{N_c} \sin ( \omega \cdot ( j + l) \cdot \Delta t +
  \omega \cdot n \cdot \Delta t \cdot k + \phi) \sin \left( 2 \pi \left(
  \frac{4 l - n}{4 l} \right) k \right) . 
\end{eqnarray}



\end{document}